# Study of anharmonicity in Zirconium Hydrides using inelastic neutron scattering and ab-initio computer modeling

Jiayong Zhang[a,b], Yongqiang Cheng[b], Alexander I. Kolesnikov[b], J. Bernholc[a,c], Wenchang Lu[a,c], Anibal J. Ramirez-Cuesta[b]

a Department of Physics, North Carolina State University, Raleigh, NC 27695, USA

b Neutron Scattering Division, Oak Ridge National Laboratory, Oak Ridge, TN 37831, USA

c Computational Sciences and Engineering Division, Oak Ridge National Laboratory, Oak Ridge, TN 37831, USA

**Abstract**

The anharmonic phenomena in Zirconium Hydrides and Deuterides, including $\varepsilon$-ZrH$_2$, $\gamma$-ZrH, and $\gamma$-ZrD, have been investigated from aspects of inelastic neutron scattering (INS) and lattice dynamics calculations within the framework of density functional theory (DFT). The observed multiple sharp peaks below harmonic multi-phonon bands in the experimental spectra of all three materials did not show up in the simulated INS spectra based on the harmonic approximation, indicating the existence of strong anharmonicity in those materials and the necessity of further explanations. We present a detailed study on the anharmonicity of zirconium hydrides/deuterides by exploring the 2D potential energy surface of hydrogen/deuterium atoms, and solving the corresponding 2D single-particle Schrodinger equation to get the eigenfrequencies. The obtained results well describe the experimental INS spectra and show harmonic behavior in the fundamental modes and strong anharmonicity at higher energies.

## I.  INTRODUCTION

Metal hydrides have been studied extensively during the past decades. In particular, there are special interests for their properties and applications in various areas, including hydrogen storage, utilization in nuclear reactors as neutron moderator [1] and fuel rod cladding materials [2] due to their low thermal neutron absorption cross-section and good mechanical properties, and also in the design of new types of actinide hydride fuels [3]. Therefore, numerous experimental and theoretical investigations on their structures and phonon properties have been conducted and reported in publications since the 1950s.

It is well known that zirconium hydrides have a complex phase diagram as a function of temperature, hydrogen content, and pressure [4–6]. Different phases such as ζ-ZrH$_{0.5}$ [7], γ-ZrH, δ-ZrH$_{1.5}$, ε-ZrH$_2$ have been identified. A stoichiometric γ-ZrH (face-centered orthorhombic metal sublattice) can be prepared by special temperature treatment [6]. Of those phases, this study focuses on γ-ZrH in Cccm space group, and ε-ZrH$_2$ in I4/mmm space group. Though the phase diagram has been determined, some hydrides' structures, stabilities, or formation mechanisms are still unknown [8].

The optical phonon properties of zirconium hydride were first analyzed using inelastic neutron scattering (INS) by Andresen *et al.* in 1957 [9]. Yamanaka [10] studied the thermal and mechanical properties of zirconium hydrides. Based on first-principles calculations, mechanical properties and thermodynamical properties of zirconium hydrides have also been broadly studied. Early studies by Ackland found that for ε-ZrH$_2$, bistable crystal structures exist with minimal energy difference [11]. The subsequent studies found that one of the bistable structures may be unstable or metastable. Elsasser found that for the center of the Brillouin zone (Gamma point) of ZrH$_2$, the potential energy profiles for hydrogen is a parabola up to about 300 meV and deviates at higher energies [12]. Besides theoretical studies, INS spectroscopy has also been employed to measure the vibrational spectra of ε-ZrH$_2$. The split peaks of multi-phonon events were observed in 1983 by Ikeda [13], yet was not explained. Kolesnikov *et al.* later attributed those split peaks below multi-phonon events in γ-ZrH and γ-ZrD to the bound multi-phonon states [14,15].

Those experiments and calculations revealed that hydrogen vibrations in zirconium hydrides are highly anharmonic. Though phonon theory and techniques have been well developed, most of them are based on harmonic situations with a limited focus beyond that point. Phonon density of states (PDOS) with full consideration of anharmonicity can be extracted from molecular dynamics (MD) simulations by the velocity autocorrelation function (VACF); however, only frequencies information can be obtained in this case, which is insufficient for further phonon analysis. Recently, the temperature-dependent effective potential (TDEP) method has been developed to address the anharmonicity problem by the MD method and taking temperature effects into consideration [16,17]. However, these approaches are incapable of building directly comparable results with our broadband INS data for zirconium hydrides. Thus, an alternative technique is needed to address the problem. Besides, with a simple structure and large volume of experimental and simulation data, zirconium hydrides are excellent candidates to directly and unambiguously assess the quality of our anharmonic analysis.

Based on the considerations above, we present a qualitative and semiquantitative analysis that accounts for the effects observed in the INS spectra of zirconium hydrides by systematically investigating the potential energy surface (PES) of these materials using the density functional theory (DFT) and quantum theory. We sample PES in 2D planes due to considerations of the balance of practicality and accuracy. In this work, we successfully combined DFT and direct solutions of Schrodinger equations with INS spectra calculations. This work will hopefully improve our understanding of the mechanical and thermal properties of zirconium hydrides and can be applied to other metal hydride systems. The isotope effects were also investigated by checking the γ-ZrD system.

## II. METHODS

Multiple *ab initio* simulation packages within the framework of DFT have been employed to carry out first-principles calculations in this work, including the Vienna *ab initio* simulation package (VASP) [18,19], and the Real-space Multi-Grid code (RMG) [20–22]. VASP is a plane-wave based DFT package with the implementation of multiple pseudopotentials. In all our VASP calculations, the electron-ion interaction and exchange-correlation functional were described by the projector-augmented wave (PAW) method [23] and the generalized gradient approximation (GGA) [24] with the parametrization of Perdew-Burke-Ernzerhof (PBE) [25], respectively. First-order Methfessel-Paxton scheme [26] with a smearing width of 0.05 eV was employed to integrate total energy in the Brillouin Zone (BZ), and the energy cutoff in the plane-wave functions was set to be 600 eV. Valence electron configuration in zirconium was carefully chosen as $4s^2 4p^6 4d^2 5s^2$ to include semi-core states, and in hydrogen was $1s^1$. The self-consistent convergence of the total energy calculation is $10^{-8}$ eV. RMG is an electronic structure calculation code based on the real-space method, implementing ultrasoft pseudopotential and norm-conserving pseudopotential. The RMG phonon methodology has been recently developed [27], and it is well suited for large-scale phonon calculations. In all RMG calculations, the real-space grid spacing was set to 0.15 Å, and norm-conserving pseudopotential with exchange-correlation described by PBE functional was used. Valence electrons for Zr were $4d^25s^2$. The convergence criteria for the root mean squared (RMS) change in the total potential energy per step was set as $10^{-7}$, and a Fermi Dirac occupation type with electron temperature as 0.05 eV was used in energy integration.

The phonon dispersions of $ZrH_x$ ($ZrH_x$ means one of ε-$ZrH_2$, γ-ZrH, and γ-ZrD) were calculated using the Phonopy package [28] employing the forces calculated by both VASP and RMG packages. Phonopy is an open-source package with interfaces to many commonly used DFT codes, including VASP, with built-in features such as phonon calculations, thermal physics calculations, etc. The maximum residual force in geometry optimization was 1 meV/Å. 2×2×2 supercells of $ZrH_x$ containing 96 (ε-$ZrH_2$) or 64 (γ-ZrH and γ-ZrD) atoms have been used for phonon calculations in the finite displacement method (FDM) with small atomic displacements of 0.01 Å in supercells. A Gamma centered 5×5×5 k-mesh for ε-$ZrH_2$, 11×11×11 for γ-ZrH and γ-ZrD was used to integrate total energy. K-mesh and energy cutoff have been tested to be converged with respect to total energy ranging up to 1 meV/atom in RMG and VASP. The phonon properties computed by the Phonopy package was used to simulate the INS spectra [29] by the OCLIMAX software [30,31], a program aiming at simulating INS spectra based on vibrational normal modes with the capability of including coherent effects and temperature effects. A 31×31×31 uniformly sampled k-mesh in the BZ was used to represent vibrational modes for our OCLIMAX calculations.

The frozen phonon calculations and potential surface sampling calculations were done with the VASP package.

To validate simulation results, experimental INS spectra for ε-$ZrH_2$ have been measured at the VISION spectrometer [32] at Spallation Neutron Source (SNS) in the Oak Ridge National Laboratory (ORNL), and INS spectra for γ-ZrH and γ-ZrD were measured previously [14,15] at the TFXA spectrometer [33] at ISIS in the Rutherford Appleton Laboratory, UK. INS can measure the intensity of neutrons' direct interaction with nuclei with respect to both energy transfer and momentum transfer, with advantages of broad bandwidth, no selection rules, nondestructive probe, and more importantly, INS simulations are straightforward. Thus, INS becomes a perfect technique to be combined with theoretical analysis.

### III. RESULTS AND DISCUSSION

A. Atomic structures and phonon properties of ZrH$_x$

The crystal unit cells of ZrH$_x$ and their phonon properties are shown in Fig. 1. The lattice vectors (a, b, c) used in our simulations for ε-ZrH$_2$ [34], γ-ZrH(γ-ZrD) [14] are (4.97728, 4.97728, 4.449) and (4.549, 4.618, 4.965) in Å, respectively. Note that phonon properties (especially optical modes) are sensitive to lattice parameters; thus, no geometry optimization is performed on the lattice vectors of these three materials to better compare to experiments. The phonon dispersions and phonon density of states (PDOS) for all three samples were calculated by DFT. Note that large phonon band gaps exist between high energy modes and low energy modes for all materials.

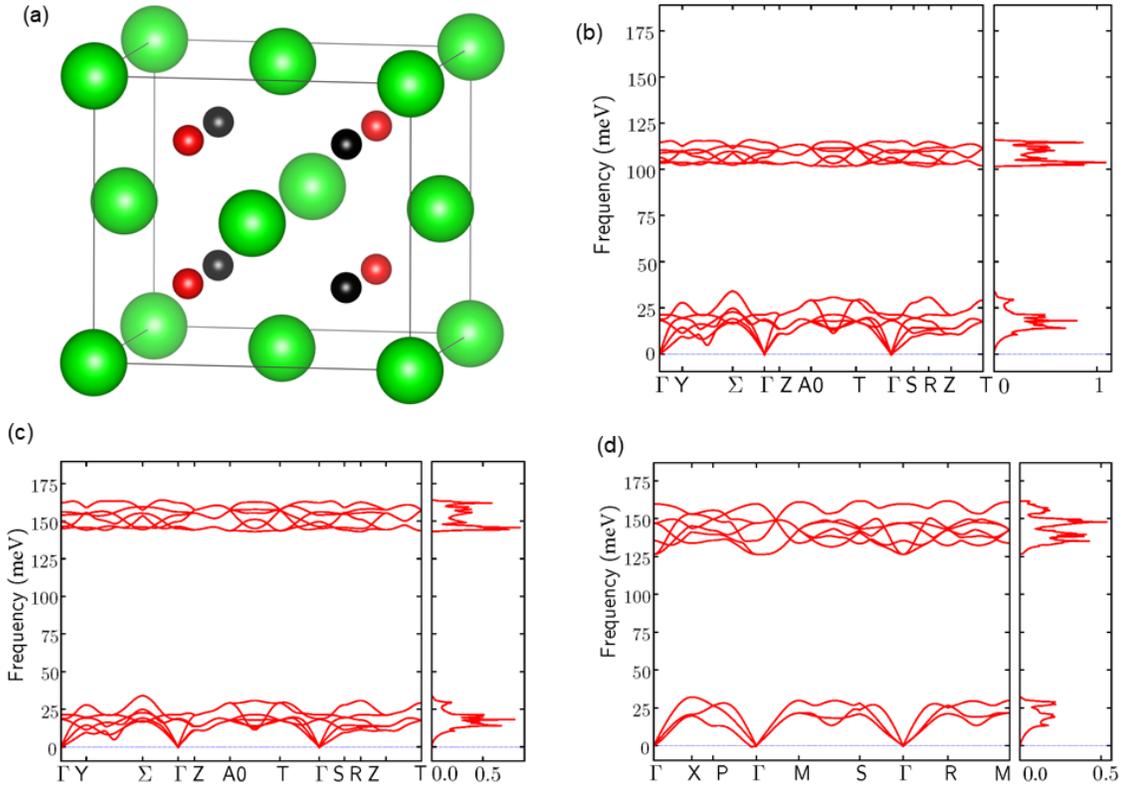

Fig. 1. (a) Crystal unit cells for ε-ZrH$_2$ (space group I4/mmm), γ-ZrH and γ-ZrD (both are of space group Cccm), where H atoms in red exist in all structures, and H atoms in black only exist in ε-ZrH$_2$. Note that in calculations, lattice constants of ε-ZrH$_2$ and γ-ZrD(H) are different. Color scheme: Zr, green; H, red/black. Figures are drawn using the VESTA program [35]. (b), (c) and (d) are calculated phonon dispersions and PDOS for γ-ZrD, γ-ZrH and ε-ZrH$_2$, respectively. The BZ paths and notation are adopted from [36].

Fig. 2 shows the experimental INS spectra for ZrH$_x$, and Fig. 3 shows simulated INS spectra for ε-ZrH$_2$ with RMG/VASP+OCLIMAX along with VISION data. In Fig. 2, within the frequency region under 1000 meV, multiple peaks representing phonon excitations are observed in all materials. Focusing on ε-ZrH$_2$, the peaks at energies lower than 30 meV correspond to lattice modes of heavy Zr atoms in the crystals. The peaks at around 150 meV are corresponding to single phonon excitations of H atoms from the ground state to their first excited states (fundamental excitations). All higher energy peaks are overtones and combinations, corresponding to multi-phonon events excited from the ground state to the states higher than the first.

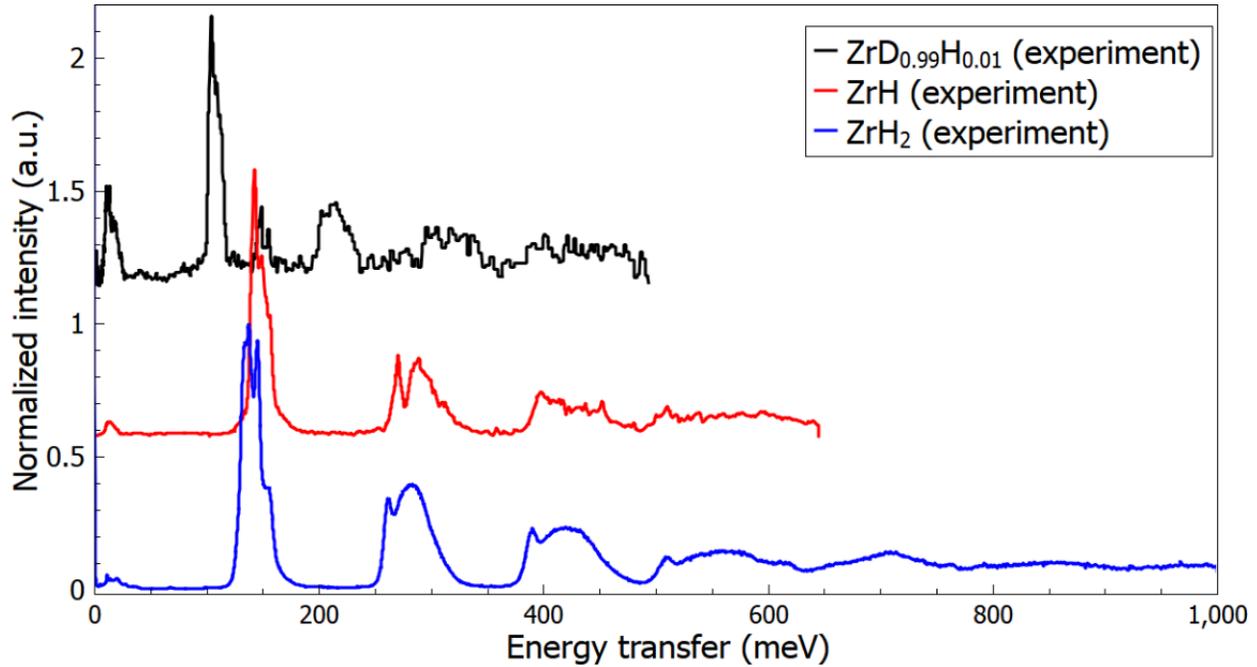

Fig. 2. The INS spectra for ε-ZrH$_2$ measured with the VISION instrument at 5K (blue), γ-ZrH (red) at 4.5K and γ-ZrD (black) at 30K measured with the TFXA spectrometer. Spectra have been normalized to the same maximum value of the fundamental peaks for comparison purposes. Note that γ-ZrD sample has been contaminated by ~1 at.% H, and also contained α- and δ-phases [14], with the actual ratio of Zr atoms in the different phases as 0.718(α-ZrD0.001)+0.269(γ-ZrD0.98)+0.013(δ-ZrD1.2), which results in extra peaks at ~140 meV and higher intensity of the lattice modes (<30 meV) compared to the expected spectrum for pure γ-ZrD. Figures are drawn using the Mantid program [37].

The INS spectra simulated from VASP and RMG for ε-ZrH$_2$ show excellent agreements with the experimental data at energies below 200 meV (see Fig. 3). However, discrepancies exist between the experimental data and the simulated INS spectra in the higher energy range. Overtone peaks at around 260 meV, 390 meV, and 510 meV exist only in the experimental data, suggesting the harmonic approximation is good for fundamental peaks but gets failing on higher energy excitations, and anharmonic effects need to be considered at higher energy. It should be noted that for γ-ZrH and γ-ZrD, similar split peaks were also observed in the experimental data (see Fig. 2).

The purpose of this work is to present a combined study regarding the anharmonicity of ZrH$_x$ by using first-principles calculations and neutron techniques.

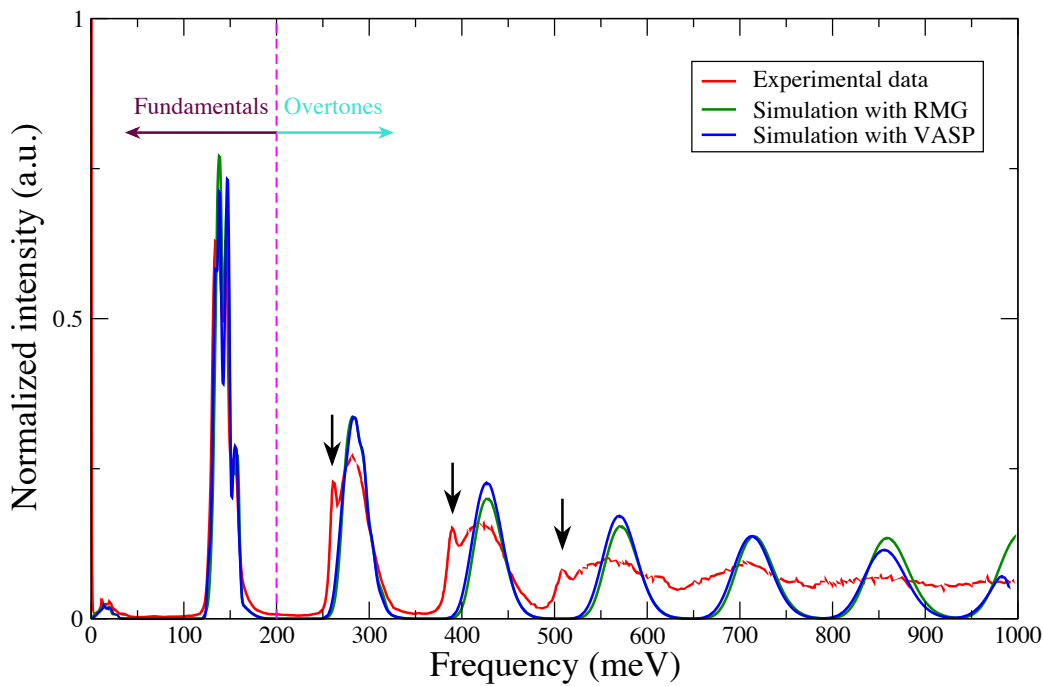

Fig. 3. VISION INS spectrum (red) at 5K and simulated spectra with RMG (green) and VASP (blue) at 0K of ε-ZrH$_2$. Spectra are normalized with respect to their area under the fundamental spectra curve. Black arrows are marking peaks resulted from anharmonic effect (split peaks). The simulated data assumes the harmonic approximation.

B. Harmonic approximation: frozen phonon method

Frozen phonon calculations were performed to explore the hydrogen potential energy profiles in ε-ZrH$_2$ as it usually captures the anharmonicity (if there is). The sampling directions of potential energy profiles were along the phonon polarization vectors, and the sampling displacements of H were chosen to be up to 1.0Å with a step of 0.1Å. Also, to reduce the computational cost, six low-frequency lattice modes (smaller than 30 meV) were not included in our frozen phonon calculations.

Our results show that all modes can be well fitted to parabolas, suggesting no anharmonic effects can be seen within displacement of 1.0 Å (which corresponds to energy range as high as few eVs) along the polarization directions, the frozen phonon method does not work in this case.

C. A direct method: mapping eigenfrequencies from Schrodinger equations with simulated INS spectra

To obtain the phonon frequencies, an intuitive and accurate way is to solve the many-body Schrodinger equation representing the vibrations of all atoms directly. The Schrodinger equation may be easily solved for single-particle; however, it becomes impossible to solve as the system size significantly increases. Thus, we will try to simplify the complex many-body problem in two steps.

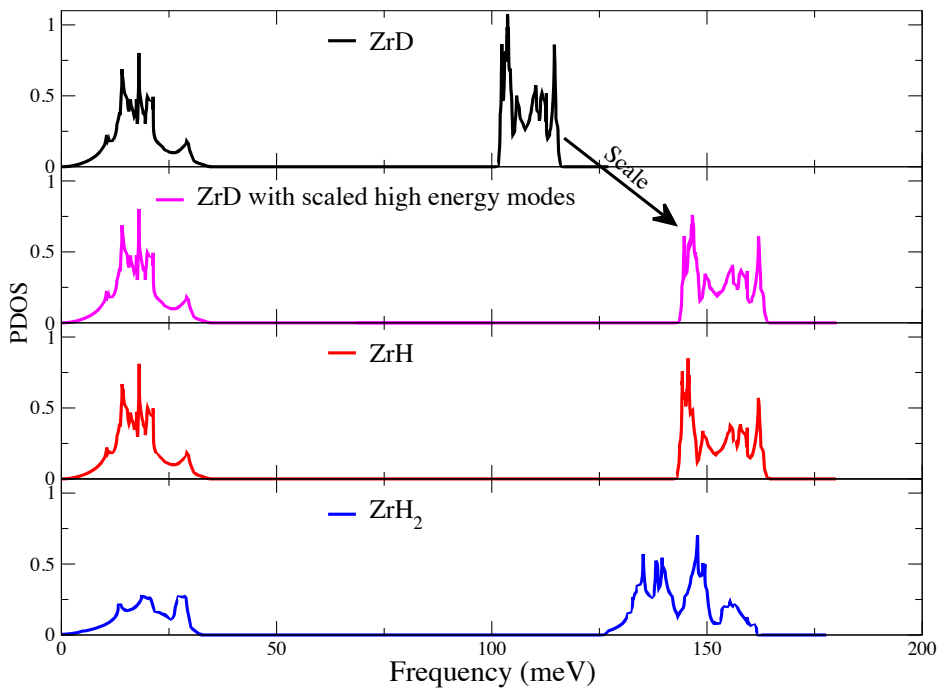

Fig. 4. PDOS comparisons between γ-ZrD (black) and scaled high energy modes of γ-ZrD (magenta), γ-ZrH (red), and ε-ZrH$_2$ (blue). The high energy modes of γ-ZrD are scaled by a factor of $\sqrt{2}$ in energy, and $1/\sqrt{2}$ in intensity to obey DOS properties, which are represented by the black arrow. Low energy modes are almost identical between γ-ZrH and γ-ZrD, while high energy modes are nearly identical after scaled.

First, separate vibrations of H/D from vibrations of Zr. The justifications of this separation come from: 1) PDOS comparison between γ-ZrH and γ-ZrD as shown in Fig. 4. The high energy modes of γ-ZrD are scaled by a factor of $\sqrt{2}$ in energy, and $1/\sqrt{2}$ in intensity. The calculated PDOS in Fig. 4 shows that the low energy modes of γ-ZrD and γ-ZrH are almost identical (the theoretical ratio is 1.005 if Zr and H/D move in-phase), while the high energy modes differ by a factor of $\sqrt{2}$ (which is close to the frequency ratio $\omega_H/\omega_D = \sqrt{m_D/m_H} = 1.4136$ if we assume the H/D and the Zr are almost independent oscillators. In the harmonic approximation, the frequency of an oscillator with mass $m$ is $\omega \propto \sqrt{k/m}$ where $k$ is the force constant and $m$ is the mass of the particle.), suggesting that Zr and H are almost independent oscillators. Olsson et al. draw similar conclusions on ε-ZrH$_2$ and ε-ZrD$_2$ [38]; 2) the charge densities of Zr and H in ε-ZrH$_2$ have near-spherical distribution with slight deformations, which indicates the Zr-H bonds and their couplings are weak [39]; 3) the mean square displacement (MSD) of Zr is much smaller than H/D's, as calculated by [29] at 0K:

$$\boldsymbol{u}_l^2 = \frac{2.09}{m_l} \sum_q w_q \sum_v \frac{|\epsilon_l(q,v)|^2}{\omega_{q,v}} \quad (1)$$

where $l$ is the atomic index, $v$ is the index of normal mode, $q$ is the index of points in BZ, $m_l$ is $l$th atom's mass in atomic units, $\omega_{q,v}$ is the vibrational frequency in the unit of meV, $\epsilon_l(q,v)$ is $l$th atom's polarization vector of $v$th mode at point $q$, and $w_q$ is the normalized weight of point $q$. Three lowest energy acoustic modes are not included in our calculations. The MSD is calculated within the low energy range (<60 meV) and the high energy range (> 60 meV), respectively. The results are listed in Tab. 1:

| Property \ Energy Range | < 60 meV | > 60 meV |
|---|---|---|
| ε-ZrH$_2$ | | |
| MSD(H) | 0.0019 (0.0436) | 0.0433 (0.2082) |
| MSD(Zr) | 0.0041 (0.0642) | 0.00000396 (0.00199) |
| $\frac{\text{MSD(H)}}{\text{MSD(Zr)}}$ | 0.46 (0.68) | 10934.34 (104.62) |
| γ-ZrH | | |
| MSD(H) | 0.0024 (0.0490) | 0.0407 (0.2019) |
| MSD(Zr) | 0.0044 (0.0663) | 0.00000236 (0.00154) |

| $\frac{\text{MSD(H)}}{\text{MSD(Zr)}}$ | 0.55 (0.74) | 17245.76 (131.10) |
|---|---|---|
| γ-ZrD | | |
| MSD(D) | 0.0024 (0.0490) | 0.0286 (0.1690) |
| MSD(Zr) | 0.0044 (0.0663) | 0.000006738 (0.0025954) |
| $\frac{\text{MSD(D)}}{\text{MSD(Zr)}}$ | 0.55 (0.74) | 4244.58 (65.12) |

Tab. 1. Calculated MSD (and √MSD in brackets) of Zr at (0, 0, 0), H/D at (¾, ¼, ¾) for ZrH$_x$ in different energy range, units are in Å² for MSD and Å for √MSD. In γ-ZrH(γ-ZrD), MSD(Zr) varies slightly at different Zr positions (within 3%).

In the low energy range less than 60 meV, Zr and H/D atoms move with comparable amplitudes. However, in the energy range greater than 60 meV, the larger ratio of $\sqrt{MSD(H)/MSD(Zr)}$ is indicating that H/D atoms are moving with larger amplitudes than Zr atoms at the same frequency.

Thus, it can be concluded that the vibrations of H/D can be viewed as fast degrees of freedom that are decoupled from the dynamics of Zr atoms.

Second, per conclusions in the first step, the Hamiltonian of a single H/D atom is:

$$\mathbf{H} = -\frac{\hbar^2}{2m}\mathbf{\nabla}^2 + V(\mathbf{r}) \quad (2)$$

where $\mathbf{\nabla}^2$ is the Laplacian operator on the H/D atom's wavefunctions, and $V(\mathbf{r})$ is the potential energy at position $\mathbf{r}$ (a set of $V(\mathbf{r})$ constitute the potential energy surface, PES). In our calculations, $V(\mathbf{r})$ are the DFT sampled potential energies of the H/D atom with substitution of its energy at the equilibrium position $\mathbf{r_0}$: $V(\mathbf{r}) = E_{DFT}(\mathbf{r}) - E_{DFT}(\mathbf{r_0})$. For the sake of simplicity and to present a simpler, yet robust explanation of the anharmonic effects, we use a 2D representation of $V(\mathbf{r})$ and thus $\mathbf{H}$. The 2D PES samplings provide a qualitative and semi-quantitative explanation of the anharmonicity in ZrH$_x$.

By checking the structure of ε-ZrH$_2$, we decided to sample the (112) plane (shifted so that the equilibrium position of the H atom is included in the plane) as this is one of the planes with the highest symmetry. The plane was sampled at a mesh density of 100 points per direction (mesh grid spacing around 0.005 Å, 10,000 mesh grids totally). Based on the sampled grid, a finer grid with a dimension of 1200×1600 = 1,920,000 was generated using a linear interpolation method to build the Hamiltonian matrix according to Eq. (2). A similar scheme has been applied to γ-ZrH and γ-ZrD. A view of the plane and the sampled PES is shown in Fig. 5.

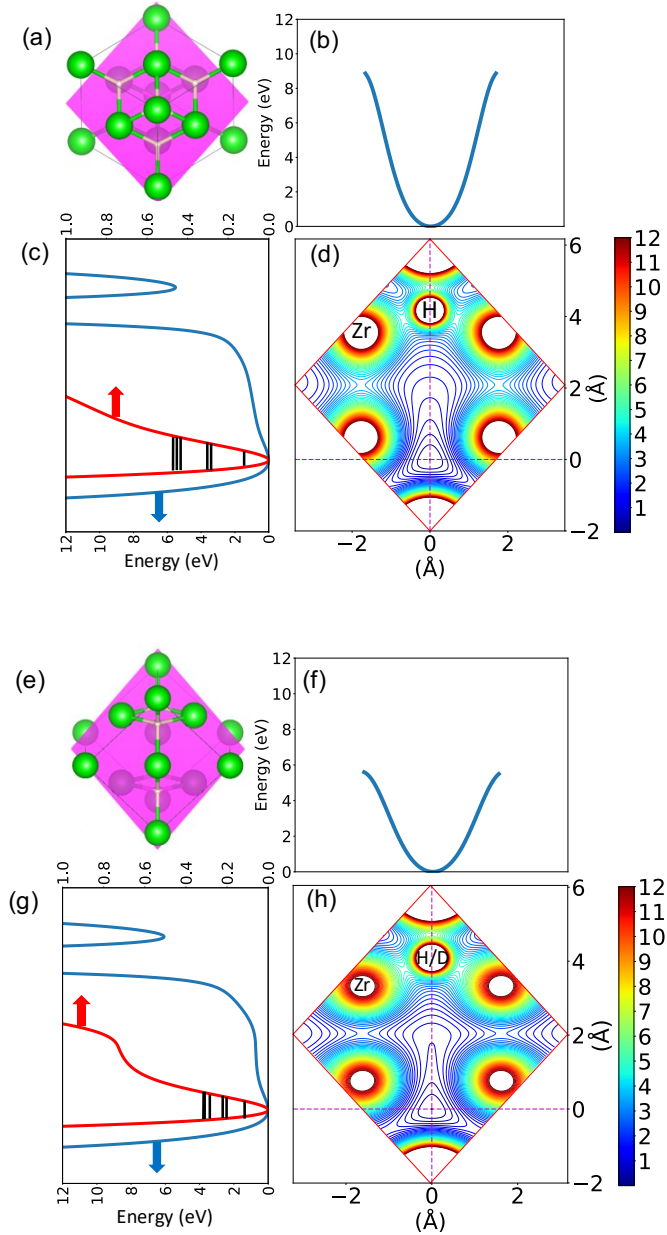

Fig. 5. Crystal unit cells with sampling plane, contour plots and sectional views (blue curves) of PES for ZrH$_x$. (a)(b)(c)(d): plane (112) of ε-ZrH$_2$; (e)(f)(g)(h): plane (-112) of γ-ZrH and γ-ZrD. Color scheme: Zr, green; H, white; sampling plane: pink. The energy range within (0, 12) eV with 50 levels are used in contour plots, with minimum energies of 0 are all located at (0, 0). Red lines in (c) and (g) are enlarged plots in vertical directions within the energy

range of (0, 1) eV, and black lines are indicating the first six eigenfrequencies (see Fig. 6) from Schrodinger equations of ε-ZrH$_2$ and γ-ZrD, respectively.

After solving the corresponding single-particle Schrodinger equation of H/D atom, the eigenfrequencies and wavefunctions which represent the energy states of atoms in the PES can be displayed, as plotted in Fig. 6. The three materials' eigenfrequencies have also been plotted in Fig. S2 [41] in the supplemental material, and direct comparisons between eigenfrequencies and experimental spectra can be found in Fig. 7. It can be seen that the anharmonic peaks in experimental spectra can be reasonably described by eigenfrequencies from solving the 2D Schrodinger equation for the first few overtones (black dashed lines), and the discrepancies become larger at high energy levels. For example, in the case of ε-ZrH$_2$, the eigenfrequencies at around 264, 384, 493 meV are below the harmonic multi-phonon bands in magenta lines, and thus indicating the existence of anharmonicity. The eigenfrequency at 264 meV agrees well with the experimental value, yet the larger eigenfrequencies at 384 and 493 meV have around 6 and 17 meV's differences with experimental values, respectively, and the discrepancies are increasing at larger eigenfrequencies. Note that in (c), the ZrD sample was contaminated with H, and thus introduced the extra low-frequency peaks, e.g., at around 140 meV.

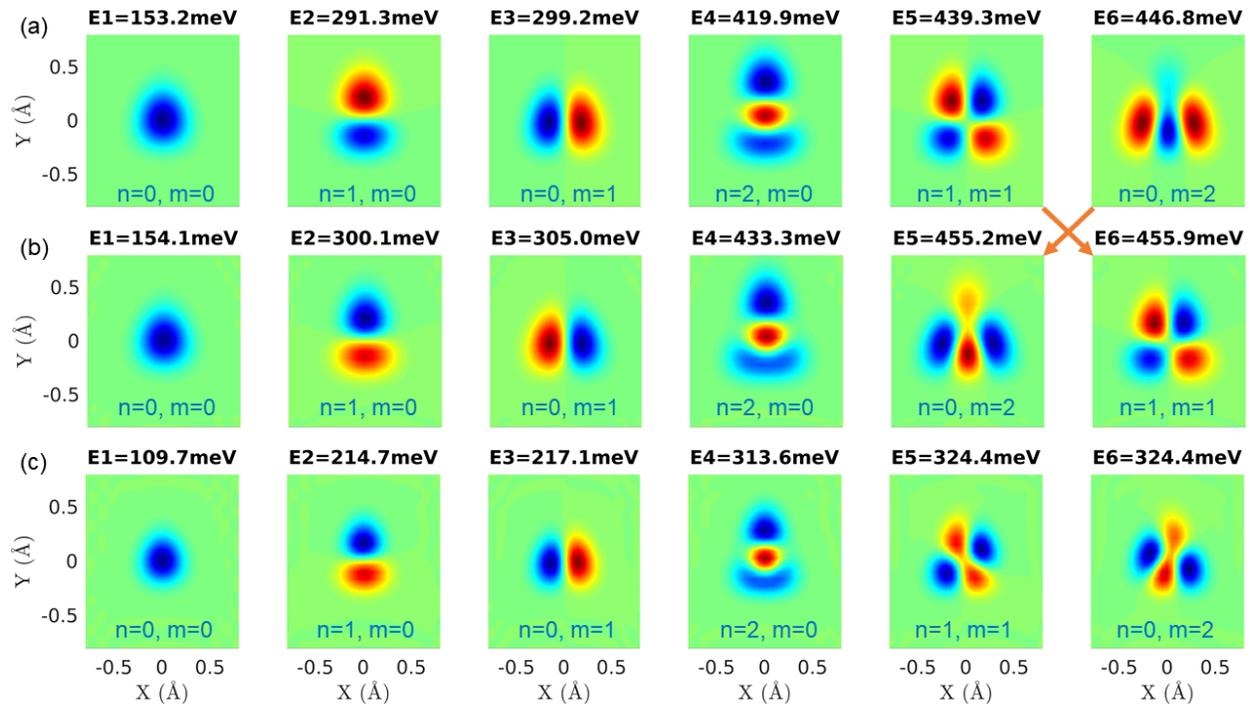

Fig. 6. First six eigenfrequencies and wavefunctions plotted in a box of range [-0.8, 0.8] Å × [-0.8, 0.8] Å by solving PES Schrodinger equations for: (a) (112) plane of ε-ZrH$_2$, (b) (-112) plane of γ-ZrH, and (c) (-112) plane of γ-ZrD. Corresponding quantum numbers are denoted in [n, m], where n is in the Y direction and m is in the X direction. Notice that cross over happens between the quantum numbers of ε-ZrH$_2$ and γ-ZrH, which are marked by orange arrows. The last two wavefunctions in γ-ZrD are degenerate. The Schrodinger equations were solved using [40].

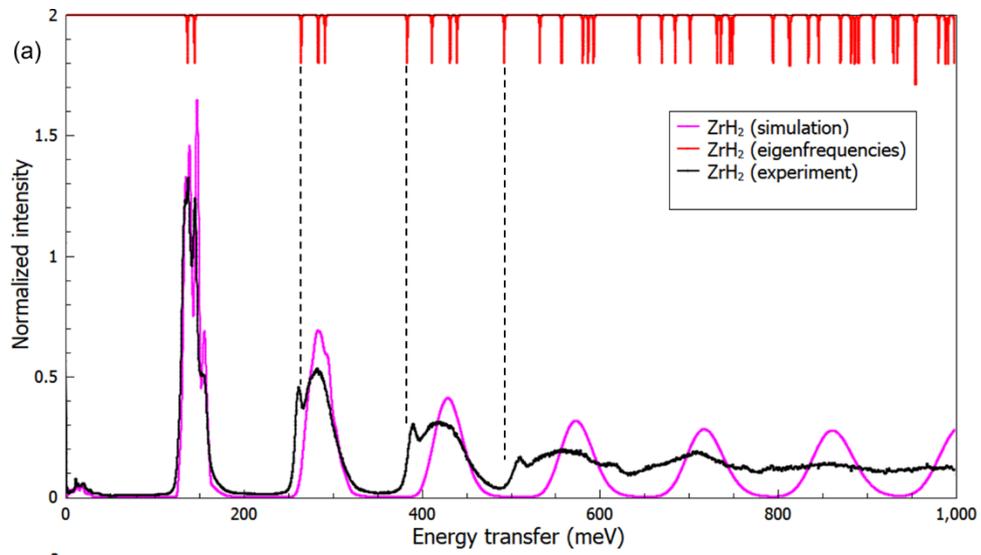
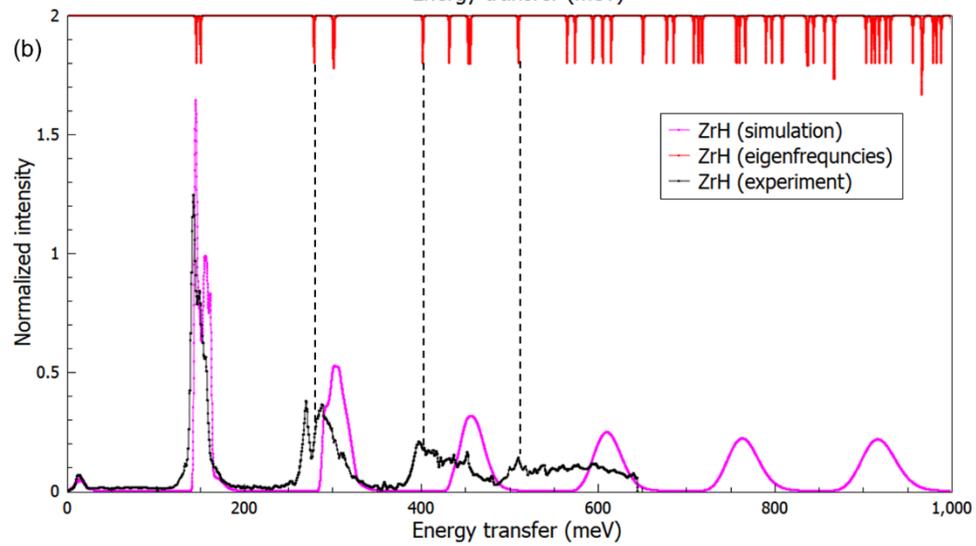
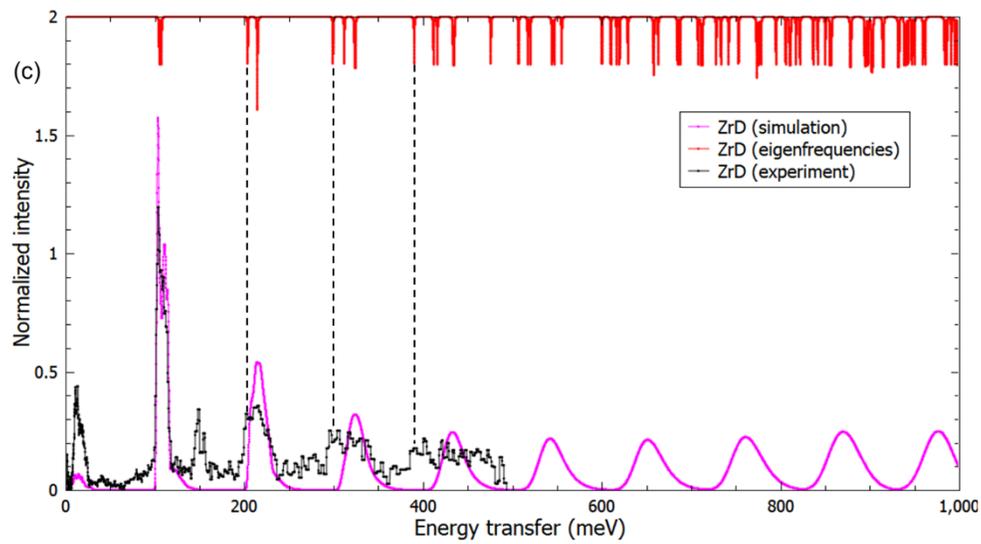

Fig. 7. Comparisons between eigenfrequencies from Schrodinger equations and experimental INS spectra for: (a) ε-ZrH$_2$, (b) γ-ZrH, (c) γ-ZrD. Eigenfrequencies from Schrodinger equations are in red lines, experimental INS spectra are in black solid lines, simulated INS spectra are in magenta lines, and the black dashed lines are for eye guidance of comparisons between eigenfrequencies and anharmonic peaks for the first to third overtones.

The eigenfrequencies from the Schrodinger equation include phonon properties in real space (PES). However, they are for the Gamma point only (BZ center) in reciprocal space, and it loses the information of phonon properties in other points. On the other hand, simulated INS spectra from DFT can sample the whole BZ (with fine enough grids) and include phonon information over the whole reciprocal space within the harmonic approximation. In order to qualitatively incorporate the anharmonicity into this model and to better compare experimental spectra and simulated spectra with the incorporated anharmonicity, we convolute the resulting anharmonic spectra at the center of the BZ from the Schrodinger equation, with the corresponding overtones calculated by DFT within the harmonic approximation. The change thus introduced to the peak positions is a valid approximation, because the eigenvalues are very sensitive to small changes in the dynamical matrix, while the eigenvectors are not that sensitive to these changes and reflect the symmetry of the system [29].

$$S_{adjusted}(n) = S_n * \{\omega_1, \omega_2 \cdots \omega_m\} = \frac{1}{m}\sum_{i=1}^{m} S_n(\omega_n - \omega_i) \quad (3)$$

$$S_{adjusted} = \sum_{n=1}^{N} S_{adjusted}(n) \quad (4)$$

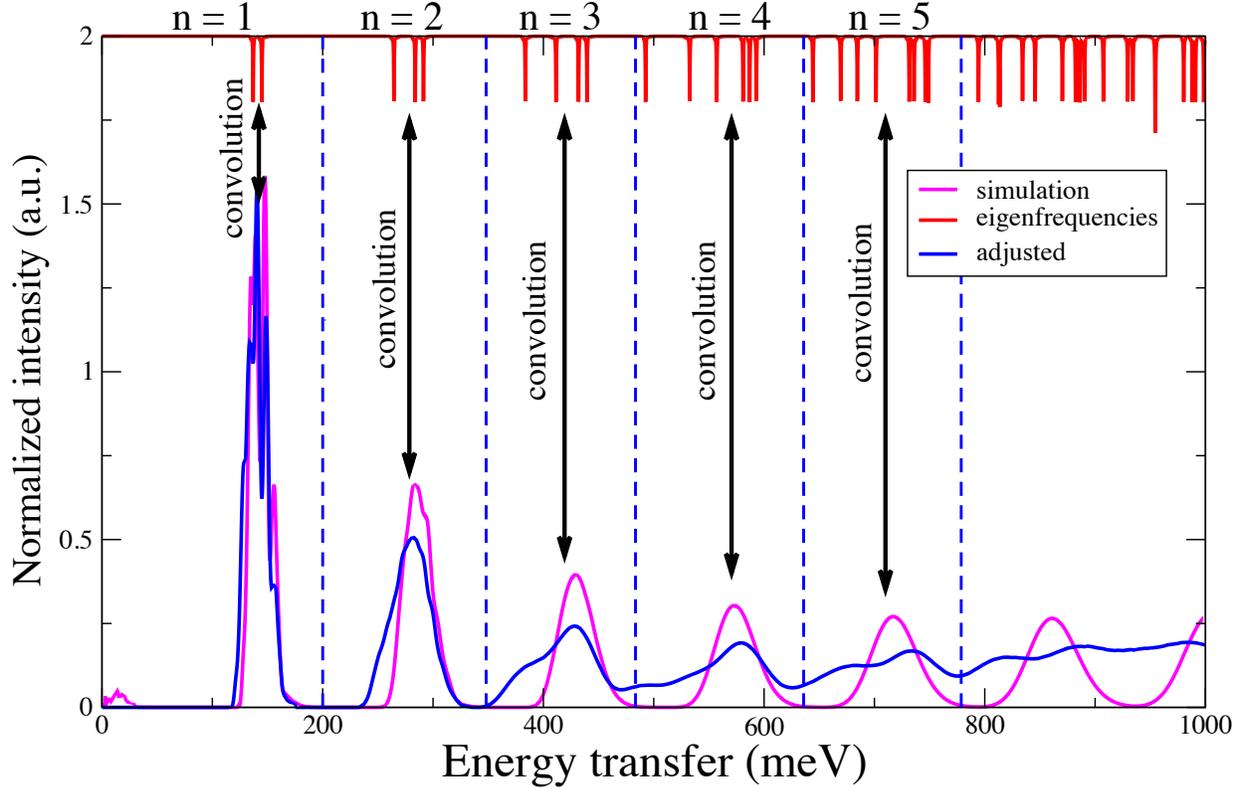

Fig. 8. A schematic figure showing the convolution procedure. The convolution procedure was done within each individual quantum event indexed by $n$. Magenta lines are indicating simulation results, red are results from direct solutions from the Schrodinger equation, and blue lines are adjusted results after convolution between simulations and eigenfrequencies.

The basic idea is, for each quantum event, to convolute the simulated spectra by DFT with eigenfrequencies from the Schrodinger equation as shown in Eq. (3), where $n$ is the index of the quantum event, $\omega_i$ ($i = 1 \ldots m$) is the $i$th eigenfrequency of the 2D Schrodinger equation for quantum event $n$, $S_n$ are the simulated spectra for quantum event $n$, $\omega_n$ is the "center of mass" (which is calculated by averaging spectra's energy transfer weighted by their intensity) of $S_n$, and $S_n(\omega_n - \omega_i)$ is a replica of $S_n$ but shifted by the difference between $\omega_n$ and $\omega_i$, $S_{adjusted}(n)$ are the final adjusted spectra to be compared with experimental data for this quantum event. The adjusted spectra for each quantum event are summarized to be the total adjusted spectra $S_{adjusted}$ as shown in Eq. (4). A schematical representation of the convolution procedure is plotted in Fig.8, and the adjusted results of ZrH$_x$ are shown in Fig. 9. It should be noted that the convolution process here does not have physical meanings, and it is mostly for comparison purposes.

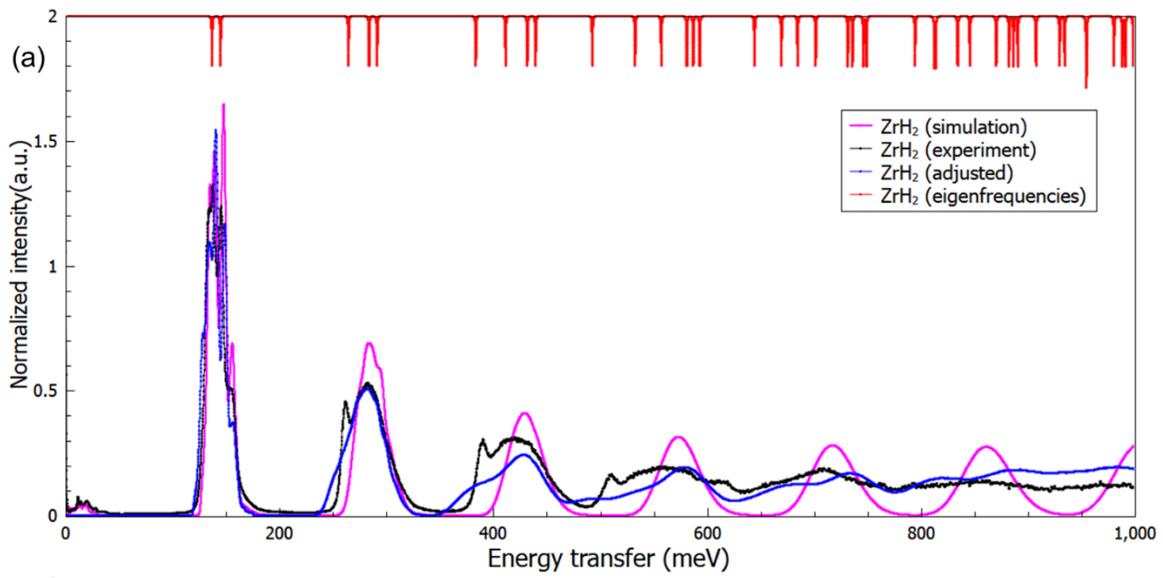
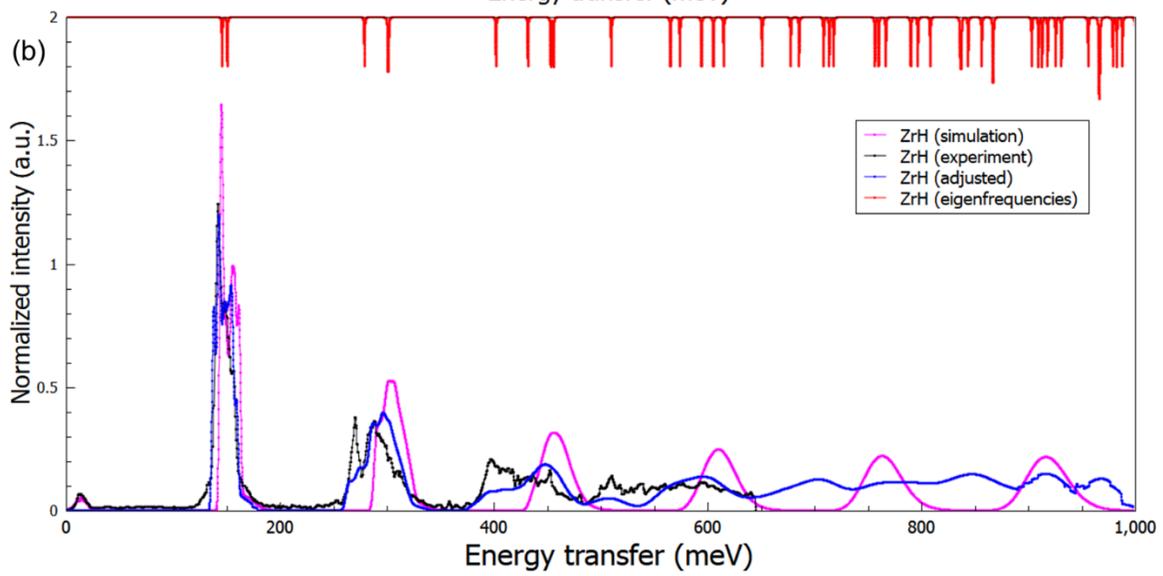
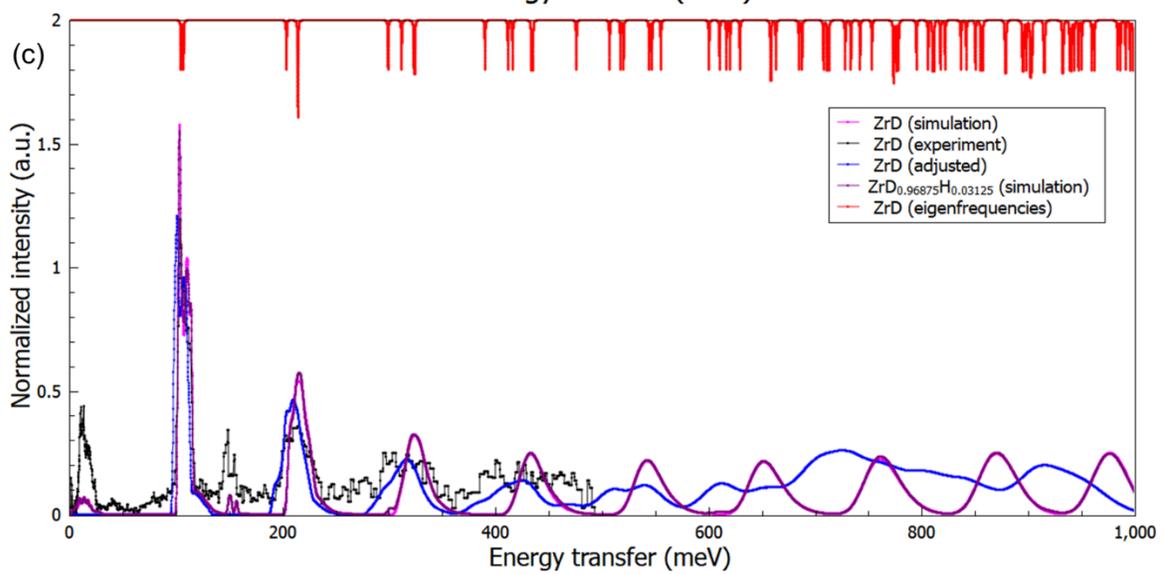

Fig. 9. INS spectra for: (a) ε-ZrH$_2$, (b) γ-ZrH, (c) γ-ZrD. Eigenfrequencies from Schrodinger equations are in red lines, experimental INS spectra are in black lines, simulated INS spectra are in magenta lines, blue lines represent adjusted spectra after convolution, and purple lines in (c) are simulated spectra of contaminated γ-ZrD. Spectra are normalized with respect to their area under the fundamental spectra curve. The ratio of neutron scattering cross-section of H and D is 82.02/7.64=10.74, and m(D)/m(H)=2, therefore fundamental modes intensities ratio of 3% H to 97% D should be ~10.74*2*0.03/0.97=0.66 (without corrections for the Debye-Waller factor). Note that in (c), the simulation spectra of ZrD and ZrD$_{0.96875}$H$_{0.03125}$ are almost on top of each other.

It can be seen that with the convolution process mentioned above, the original simulated peaks are mostly all broadened, and show better agreements with the experimental INS spectra. In the cases of γ-ZrH and γ-ZrD, we can see significant improvements in the peaks' positions, as both the original and the adjusted simulated spectra have close shifts in energy transfer on multiple peaks. Since the γ-ZrD sample was contaminated with H, H atoms of concentration 3.125 at.% (1 in 32 D atoms in the simulation box was substituted by H atom) were also added to the pure γ-ZrD in our simulations. The calculation well reproduces the splitting of the hydrogen local mode peak at ~150 meV into two peaks (double and single degenerate, according to the symmetry of the H position), and the absence of dispersion due to disorder of the H atoms in the lattice (random defects). Notice that the relative higher intensity of low energy modes in experimental γ-ZrD$_{0.99}$H$_{0.01}$ than simulated low energy modes of γ-ZrD$_{0.96875}$H$_{0.03125}$ is mostly due to contributed low energy modes by α-Zr [14].

Note, that the exact knowledge of vibrational dynamics of different Zr-H materials is important not only from the point of view fundamental physics and chemistry but also for practical applications in the nuclear field such as critical safety studies. Therefore the obtained results on dynamics of ZrH$_x$ materials in the current study, which describes the anharmonicity of the INS spectra at energies above 300 meV, can be very useful for criticality and neutron transport simulations of the reactors containing Zr-H materials at anticipated operational temperatures up to 1200 °C (at which the first overtone level in Zr-H is ~10% populated).

IV.     CONCLUSIONS

The anharmonicity phenomena in ε-ZrH$_2$, γ-ZrH, and γ-ZrD have been identified and thoroughly studied with techniques from both DFT and INS. The anharmonicity is not apparent on the fundamental vibrations, and it becomes evident at higher energies. This effect is not appreciable around room temperature. However, ZrH$_x$ has been considered as a neutron moderator material for nuclear reactors; in this case, the anharmonic effects cannot be ignored since neutron with high energies will scatter from the hydrogen atoms and will experience deviations from the scattering kernel predictions based on the harmonic approximation. While the harmonic model failed to capture the real potential energy surface and to explain the split peaks shown in experimental INS spectra, results by convoluting eigenfrequencies of the 2D Schrodinger equation with simulated INS spectra give a good description of the anharmonicity in these materials. By using the proposed convolution process in this article, one can see obvious anharmonic shifts of the overtones in the adjusted INS spectra which are in agreement with the experimental data. The method proposed here, combining DFT and OCLIMAX, is a possible way to explain anharmonicity in materials beyond harmonic phonon theories. This work could shed light on further studies on phonon calculations on the anharmonic systems like TiH$_x$, PdH$_x$, and other metal-hydrogen systems.

## V.     ACKNOWLEDGMENTS

This work was supported through Oak Ridge National Laboratory's Graduate Opportunity (Go!) Program in the Neutron Scattering Division, ORNL. Neutron scattering experiments were conducted at the VISION beamline at ORNL's Spallation Neutron Source, which is supported by the Scientific User Facilities Division, Office of Basic Energy Sciences (BES), U.S. Department of Energy (DOE), under Contract No. DE-AC0500OR22725 with UT Battelle, LLC. The computing resources were made available through the VirtuES and the ICEMAN projects, funded by Laboratory Directed Research and Development program at ORNL.